\documentclass[conference]{IEEEtran}
\IEEEoverridecommandlockouts
\usepackage{cite}
\usepackage{amsmath,amssymb,amsfonts}
\usepackage{graphicx}
\usepackage{textcomp}
\usepackage{xcolor}
\usepackage{todonotes}
\usepackage{tabularx}

\usepackage{hyperref}

\usepackage[noend]{algpseudocode}

\usepackage[ruled,boxed, vlined,linesnumbered]{algorithm2e}
\usepackage{censor}
\usepackage{subfigure}

\newcommand\fig[1]{Figure~\ref{fig:#1}}

\newcommand\sect[1]{Section~\ref{sec:#1}}

\newcommand\reqn[1]{Equation~(\ref{eqn:#1})}

\graphicspath{{./figures/}}

\newcommand{\figeps}[3][]{%
 \begin{figure}[htb]%H]
  \begin{center}
   \leavevmode
      \parbox[t]{#1}{%
        \resizebox{#1}{!}{\includegraphics{#2}}
      }
      \vspace{-0.2cm}
   \caption{#3}
   \label{fig:#2}
  \end{center}
 \end{figure}
}

\title{P4-Protect: 1+1 Path Protection for P4}

\author{\IEEEauthorblockN{Steffen Lindner, Daniel Merling, Marco H\"aberle, and Michael Menth}
\IEEEauthorblockA{\textit{\{steffen.lindner, daniel.merling, marco.haeberle, menth\}@uni-tuebingen.de} \\
\textit{University of Tuebingen, Germany}}
}

\begin{document}

\maketitle

\begin{abstract}
1+1 protection is a method to secure traffic between two nodes against failures in between. The sending node duplicates the traffic and forwards it over two disjoint paths. The receiving node assures that only a single copy of the traffic is further forwarded to its destination. In contrast to other protection schemes, this method prevents almost any packet loss in case of failures. 1+1 protection is usually applied on the optical layer, on Ethernet, or on MPLS.

In this work we propose the application of 1+1 for P4-based IP networks. We define an 1+1 protection header for that purpose. We describe the behavior of sending and receiving nodes and provide a P4-based implementation for the BMv2 software switch and the hardware switch Tofino Edgecore Wedge 100BF-32X. We illustrate how to secure traffic, e.g. individual TCP flows, on the Internet with this approach. Finally, we present performance results showing that the P4-based implementation efficiently works on the Tofino Edgecore Wedge 100BF-32X.
\end{abstract}


%\todo[inline]{Repos BMv2 \& Tofino anlegen, Tofino nur privat über Faster, BMv2 öffentlich}

\section{Introduction}\label{sec:intro}
There are various concepts to secure traffic transmission against failure of path components such as links or nodes. The fastest is 1+1 protection. A sender duplicates traffic and forwards it over disjoint paths while the receiver forwards only the first copy received for every packet. In case of a failure, any packet loss can be avoided, which makes 1+1 protection attractive for highly reliable applications. 1+1 protection is implemented in optical networks to protect an entire trunk. It is also available for MPLS \cite{Y.1703} and Ethernet \cite{G.8031}, which are carrier technologies for IP and introduce signaling complexity. 
In this paper, we leverage the P4 programming language \cite{BoDa14} to provide 1+1 protection for IP networks.  We program P4 switches such that they feature IP forwarding, the sending and receiving node behaviour of 1+1 protection which includes IP encapsulation and decapsulation. 
We call this approach P4-Protect. Targets of our implementation are the software switch BMv2 and the hardware switch Tofino Edgecore Wedge 100BF-32X. A particular challenge is the selection of the fist copy of every duplicated packet at the receiver. We provide a controller that allows to set up 1+1 protection between P4 nodes implementing P4-Protect. Furthermore, protected flows can be added using a fine-granular description based on various header fields. 
We evaluate the performance of P4-Protect on the hardware switch. We show that P4-Protect can be used with only marginal throughput degradation and we illustrate that P4-Protect can significantly reduce jitter when both paths have similar delays. 

%To that end, traffic is tunneled between two endpoints over two different paths that should be as disjoint as possible. 
%The sending node duplicates and encapsulates the traffic and the receiving node decapsulates the traffic and forwards the earliest copy received for every packet.

The paper is structured as follows. 
\sect{relatedWork} gives an overview of related work. 
\sect{concept} describes the 1+1 protection mechanism used for our implementation and extensions for its use on the general Internet.
\sect{implementation} presents a P4-based implementation including specifics for the Tofino Edgecore Wedge 100BF-32X. 
We evaluate the performance of P4-Protect on the hardware switch in \sect{results} and conclude the paper in \sect{conclusion}.

\section{Related work}\label{sec:relatedWork}
We review various resilience concepts for communication networks. Afterwards, we give examples for 1+1 protection.

\subsection{Overview}
Rerouting reorganizes the traffic forwarding to avoid failed components. This happens on a time scale of a second. Fast reroute (FRR) locally detects that a next hop is unreachable and deviates traffic to an alternative next hop \cite{RFC5286}. The detection may take a few 10s of milliseconds so that traffic loss cannot be avoided. Both rerouting and FRR do not utilize backup resources under failure-free conditions, but their reaction time suffers from failure detection delay. 1:1 protection leverages a primary/backup path concept. To switch over, the head-end node of the paths needs to be informed about a failure, which imposes additional delay. With restoration, recovery paths may be dynamically allocated so that even more time is needed to establish the restoration paths \cite[p. 31]{NetworkRecovery}. 1+1 protection duplicates traffic and sends it over two disjoint paths whereby the receiving node needs to eliminate duplicates. That method is fastest, but it requires extra capacities also under failure-free conditions. Some services can afford short network downtimes, other services greatly benefit from 1+1 protection's high reliability.

The surveys \cite{Me00}, \cite{ZhSu00}, and \cite{FuVa00} provide an overview of various protection and restoration schemes. The authors of \cite{FuVa00} discuss survivability techniques for non-WDM networks like automatic protection switching (APS) and self healing rings (SHR) as well as dynamic restoration schemes in SONET. They further describe protection methods for optical WDM networks. A comprehensive overview of protection and restoration mechanisms for optical, SONET/SDH, IP, and MPLS networks can be found in \cite{NetworkRecovery}.

SDN with inband signalling increases the need for fast and local protection against failures because the controller may no longer be reachable in case of a failure or highly loaded. In addition, with SDN new protection mechanisms can be implemented, e.g., to reduce state in the network. Examples are given in \cite{Menth18g}.

\subsection{1+1 Protection}

The ITU-T specification Y.1703 \cite{Y.1703} defines an 1+1 path protection scheme for MPLS. They add sequence numbers to packets and replicate them on disjoint paths. At the end of the paths, duplicate packets are identified by the attached sequence number and eliminated. P4-Protect works similarly. 
However, it does not require MPLS. It is compatible with IP, and works over the Internet. 

The authors of \cite{OvBi12} compare several implementation strategies of 1+1 protection, i.e, traditional 1+1 path protection, network redundancy 1+1 path protection (diversity coding) \cite{AyCh93}, and network-coded 1+1 path protection.
Their analytical results show that diversity coding and network coding can be more cost-efficient, i.e., they require about 5-20\% less reserved bandwidth.
The delay impact of 1+1 path protection in MPLS networks has been investigated in \cite{NiCi10}.
McGettrick et. al \cite{McGu13} consider 10 Gb/s symmetric LR-PON. They reveal switch-over times to a backup OLT of less than 4 ms.
Multicast traffic has often realtime requirements. 
Mohandespour et. al extend the idea of unicast 1+1 protection to protect multicast connections \cite{MoKa15}.
They formulate the problem of minimum cost multicast 1+1 protection as a 2-connectivity problem and propose heuristics.
%Simulations show, that the average cost based on their best heuristic is 2.6\% higher than the optimal solution.
Braun et. al \cite{Menth17a} propose maximally redundant trees for 1+1 protection in BIER, a stateless multicast transport mechanism. It leverages the concept of multicast-only FRR \cite{RFC7431}.

\section{P4-Protect: Concept}\label{sec:concept}
We first give an overview of P4-Protect. 
We present its protection header, the protection connection context, and the operation of the protection tunnel ingress and egress nodes (PTIs/PTEs). 

\subsection{Overview}
With P4-Protect, a protection connection is established between two P4 switches. Protected traffic is duplicated by a protection tunnel ingress (PTI) node and simultaneously carried through two protection tunnels to a protection tunnel egress (PTE) node. The PTE receives the duplicated traffic and forwards the first copy received for every packet.

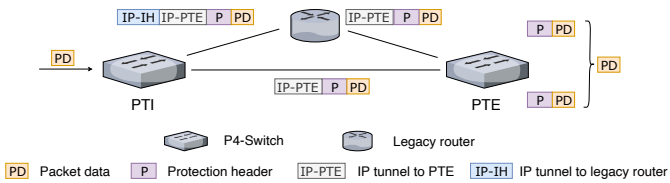
\figeps[\columnwidth]{protection_mechanism}{With P4-Protect, a PTI encapsulates and duplicates packets, and sends them over disjoint paths; the PTE decapsulates the packets and forwards only the first packet copy.}

\fig{protection_mechanism} illustrates the protocol stack used with P4-Protect. The PTI adds to each packet received for a protected flow a protection header (P) that contains a sequence number which is incremented for each protected packet. The packet is equipped with an additional IP header (IP-PTE) with the PTE's IP address as destination. The PTI duplicates that packet and forwards the two copies over different paths. The paths may be different due to traffic engineering (TE) capabilities of the network or path diversity may be achieved through an additional intermediate hop. When the PTE receives a packet, it removes its outer IP header (IP-PTE). If the sequence number in the protection header is larger than the last sequence number received for this connection, it removes the protection header and forwards the packet; otherwise, the packet is dropped. The latter is needed as duplicate packets are also considered harmful.

\subsection{Protection Header}\label{sub:ProtectionHeader}
%\fig{protection_header} illustrates the structure of the protection header. 
The protection header contains a 24 bit \textit{connection identifier} (CID), a 32 bit \textit{sequence number} (SN) field, and an 8 bit \textit{next protocol} field. 
%
%We explain their use in the following.
%\figeps[\columnwidth]{protection_header}{Structure of the protection header.}
The CID is used to uniquely identify a protection connection at the  PTE. The sequence number is used at the PTE to identify duplicates. The \textit{next protocol} field facilitates the parsing of the next header. We reuse the IP protocol numbers for this purpose.

\subsection{Protection Connection Context}
A protection connection is set up between a PTI and PTE. Their IP addresses are associated with this connection, including two interfaces over which duplicate packets are forwarded. For each connection, the PTI has a sequence number counter $SN^{PTI}_{last}$ which is incremented for each packet forwarded over the respective protection connection. Likewise, the PTE has a variable $SN^{PTE}_{last}$ which records the highest sequence number received for the respective protection connection. A CID is used to identify a connection at the PTE. A PTI may have several protection connections with the same CID but different PTEs  (see \sect{ProtConnConf}).  

\subsection{PTI Operation}\label{sec:PTI}
The PTI has a set of flow descriptors that are mapped to protection connections. If the PTI receives a packet which is matched by a specific flow descriptor, the PTI processes the packet using the corresponding protection connection. That is, it increments the $SN^{PTI}_{last}$, adds a protection header with CID, \textit{next protocol} set to \texttt{IPv4}, and the $SN$ set to $SN^{PTI}_{last}$. Then, an IP header is added using the PTI's IP address as source and the  IP address of the PTE associated with the protection connection as destination. The packet is duplicated and forwarded over the two paths associated with the protection connection.

\subsection{PTE Operation}\label{sec:PTE}
During failure-free operation, the PTE receives duplicate packets via two protection tunnels. When the PTE receives a packet, it decapsulates the outer IP header. It uses the CID in the protection header to identify the protection connection and the corresponding $SN^{PTE}_{last}$. If the $SN$ in the protection header is larger than $SN^{PTE}_{last}$, $SN^{PTE}_{last}$ is updated by $SN$, the protection header is decapsulated, and the original packet is forwarded; otherwise, the packet is dropped.

The presented behavior works for unlimited sequence numbers. The limited size of the sequence number space makes the acceptance decision for a packet more complex. Then, a $SN$ larger than $SN^{PTE}_{last}$ may indicate a copy of a new packet, but it may also result from a very old packet. To solve this problem, we adopt the use of an acceptance window as proposed in \cite{Y.1703}. The window is $W$ sequence numbers large. Let $SN_{max}$ be the maximum sequence number. If  $SN^{PTE}_{last} + W < SN_{max}$ holds, a new sequence number $SN$ is accepted if the following inequality holds:
\begin{eqnarray}
    SN^{PTE}_{last} < & SN & \leq SN^{PTE}_{last} + W
    \label{eqn:acceptance0}
\end{eqnarray}
If  $SN^{PTE}_{last} + W\geq SN_{max}$ holds, a new sequence number $SN$ is accepted if one of the two following inequalities holds:
\begin{eqnarray}
    SN^{PTE}_{last} &<& SN\\  
    SN & < & SN^{PTE}_{last} + W - SN_{max}
    \label{eqn:acceptance1}
\end{eqnarray}
This allows a packet copy to arrive  $SN_{max} - W$ sequence numbers later than the corresponding first packet copy without being recognized as new packets. 
%Thus, both protection tunnels may have a maximum difference in one-way delay of $D_{max}=\frac{SN_{max} - W}{C}$ given an identical path capacity of $C$. If a packet arrives later than $D_{max}$, it is erroneously recognized as a new packet. This suggests to choose a small $W$. However, packets may be lost on both paths. If $W$ or more packets are lost on both protection paths, the PTE erroneously recognizes new packets as old packets and drops them, which is also undesirable. Therefore, and to simplify the implementation, we choose a window size of $W=\frac{SN_{max}}{2}$. This supports a maximum delay difference of 1.6 s when minimum size packets of 40 bytes are sent with a transmission speed of $C=1$ Tb/s.
%\input{sections/p4-foundation}
\section{Implementation}\label{sec:implementation}
In this section we present the implementation of P4-Protect. We describe the supported header stacks, explain the control blocks, their organization in ingress and egress control flow, and we reveal implementation details about some control blocks. Finally, we sketch most relevant aspects of the P4-Protect controller.

\subsection{Supported Header Stacks}
Incoming packets are parsed so that their header values can be accessed within the P4 pipeline. To that end, we define the following supported header stacks. Unprotected IP traffic has the structure IP/TP, i.e., IP header and some transport header (TCP/UDP), and protected IP traffic has the structure IP/P/IP/TP, i.e., the IP header with the PTE's address, the protection header, the original IP header, and a transport header. IP traffic without transport header is parsed only up to the IP header. 

\subsection{Control Blocks}
We present three control blocks of our implementation of  P4-Protect. They consider the packet processing by PTI and PTE.

\subsubsection{Control Block: Protect\&Forward}
When the PTI receives an IP packet, it is parsed and matched against the Match+action (MAT) table ProtectedFlows. In case of a match, the packet is equipped with an appropriate header stack, duplicated, and sent to appropriate egress ports. In case of a miss, the packet is processed by a standard IPv4 forwarding procedure.

\subsubsection{Control Block: Decaps-IP}
When the PTE receives an IP packet with the PTE's own IP address, the IP header is decapsulated. If the next protocol indicates a protection header, the packet is handed over to the Decaps-P control block; otherwise, the packet is processed by the Protect\&Forward control block since the resulting packet may need to be protected and forwarded.

\subsubsection{Control Block: Decaps-P}
In the Decaps-P control block, the PTE examines the protection header and decides whether to keep or drop the packet as it is a copy of an earlier received packet. To keep the packet, the protection header is decapsulated.

\subsection{Ingress and Egress Control Flow}
The interdependencies between the control blocks suggest the following ingress control flow: Decaps-IP, Decaps-P, Protect\&Forward. At a mere PTI, no action is performed by the Decaps-IP and Decap-P control block. The Protect\&Forward takes care that protected traffic is duplicated and sent over two different paths and that unprotected traffic is forwarded by normal IPv4 operation. At a mere PTE, protected traffic is decapsulated and selected before being forwarded by normal IPv4 operation. Unprotected traffic is just forwarded by normal IPv4 operation.

\subsection{Control Block Implementations}
In the following, we explain implementation details of the Protect\&Forward control block and the Decaps-P control block. We omit the Decaps-IP control block as it is rather simple.

\subsubsection{Protect\&Forward Control Block}
The operation of the Protect\&Forward control block is illustrated in \fig{ip_control}. It utilizes the MAT ProtectedFlows to process all packets. It effects that protected traffic is encapsulated at the PTI with a protection header and an IP header for tunneling.

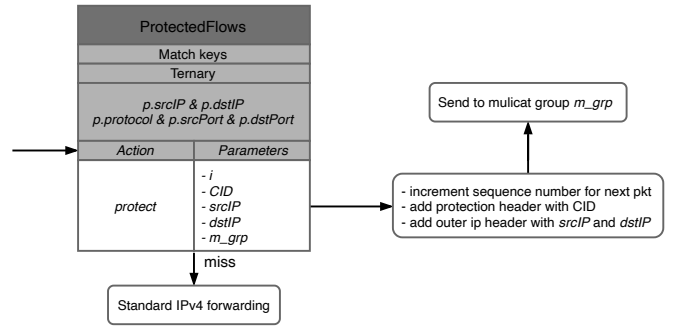
\figeps[\columnwidth]{ip_control}{The MAT ProtectedFows inside the Protect\&Forward control block is applied to IPv4 traffic.}

The MAT ProtectedFlows uses a ternary match on the classic 5-tuple description of a flow: the source and destination IP address and port as well as the protocol field. In case of a match, the MAT maps a packet to a specifc protection connection and calls the protect action with the connection-specific parameters $i$, $CID$, $srcIP$, $dstIP$, and $m\_grp$. The protect action increments the register $SN_{last}^{PTI}[i]$ where $i$ is a connection-specific index to access a register containing the last sequence number. On the Tofino target, this is performed by a separate register action. The protect action further pushes a protection header on the packet including $CID$, i.e., the connection-specific CID, $SN_{last}^{PTI}[i]$, and the next protocol set to \texttt{IPv4}. Then, it pushes an IPv4 header with the IP address $srcIP$ of the PTI as source IP and the IP address $dstIP$ of the PTE as destination IP. The protocol field of this outer IP header is set to \texttt{P4-Protect}. Finally, the multicast group of the packet is set to $m\_grp$. It is a connection-specific multicast group. It effects that the packet is duplicated and sent to two egress ports in order to deliver it via two protection tunnels to the PTE. 
In case of a miss, the packet is unprotected and handled by a standard IPv4 forwarding procedure, which is not further explained in this paper.

\subsubsection{Decaps-P Control Block}
The Decaps-P control block decides whether a packet is new and should be forwarded or dropped. It compares the sequence number $SN$ of the packet's protection header with the last sequence number of the corresponding protection connection. The latter can be accessed by the register $SN_{last}^{PTE}[CID]$ where CID is given in the protection header. The acceptance is decided based on \reqn{acceptance0} or \reqn{acceptance1} depending on the value of $SN$ and $W$ where $W$ is given as a constant. 

As the check is rather complex, it requires careful implementation for the Tofino target \footnote{Tofino is a high-performance chip which operates at 100 Gb/s so that only a limited set of operations can be performed for each packet, in particular in connection with register access.}. It leverages the fact that we set $W=\frac{SN_{max}}{2}$. Furthermore, it requires a reformulation of \reqn{acceptance0} and \reqn{acceptance1}.

If $W\leq SN$ holds, the following two inequalities must be met:
\begin{eqnarray}
    SN^{PTE}_{last} & < & SN\\
    SN - SN^{PTE}_{last} &\le& W
\end{eqnarray}
Otherwise, if $SN < W$, it is sufficient that only one of the following two inequalities holds:
\begin{eqnarray}
    SN^{PTE}_{last} & < & SN\\
    W &\le& SN^{PTE}_{last} - SN
\end{eqnarray}
Both cases are implemented as separate register actions on the Tofino target.
With 32 bit sequence numbers, a minimum packet size of 40 bytes and a transmission speed of $C = 1$ Tb/s, a delay difference up to 1.6s can be compensated.

The BMv2 version of the implementation can be found at Github\footnote{Repository: \url{https://github.com/uni-tue-kn/p4-protect}}.
The Tofino version of the implementation can be accesses as member of the \textit{Barefoot Faster Program} at Barefoot Faster\footnote{Link: \url{https://forum.barefootnetworks.com/}}.

\subsection{Controller for P4-Protect}
\label{sec:control_plane}

P4-Protect's controller offers an interface for the management of protection connections and protected flows. It configures in particular the MAT ProtectedFlows but also other MATs needed for standard IPv4 forwarding or IP decapsulation. In the following, we explain the configuration of protection connections and protected flows.

\subsubsection{Configuration of Protection Connections}\label{sec:ProtConnConf}
A protection connection is established by choosing registers on PTI and PTE  to record the last sequence numbers $SN_{last}^{PTI}$ and $SN_{last}^{PTE}$ of a protection connection. The connection identifier is the PTE's index to access $SN_{last}^{PTE}$. On the PTI, a different index $i$ may be chosen to access $SN_{last}^{PTI}$. Furthermore, the registers are initialized with zero. Moreover, the controller sets up a multicast group $m\_grp$ for each connection so that its traffic will be replicated in an efficient way to the two desired interfaces.

\subsubsection{Configuration of Protected Flows}
A protected flow is established by adding a new flow rule in the MAT ProtectedFlows of the PTI. It contains an appropriate flow descriptor and the parameters to call the action protect. Those are the index $i$ associated with the corresponding protection connection, the CID needed at the PTE to identify the protection connection, the IP address of the PTI, the IP of the PTE, and the multicast group $m\_grp$.
\section{Evaluation}\label{sec:results}
In this section we evaluate the performance of the implemented mechanism on the Tofino Edgecore Wedge 100BF-32X. First, we compare packet processing times with and without P4-Protect. Then, we demonstrate that very high data rates can be achieved with and without P4-Protect on a 100 Gb/s interface. Finally, we show that P4-Protect can provide a transmission service with reduced jitter compared to the jitter of both protection tunnels.

\subsection{Packet Processing Time}\label{sec:ProcessingTime}
P4-Protect induces forwarding complexity. To evaluate its impact, we leverage P4 metadata to calculate the time a packet takes from the beginning of the ingress pipeline to the beginning of the egress pipeline. This is sufficient for a comparison as all work for P4-Protect is done in the ingress pipeline and all considered forwarding schemes utilizes the same egress pipeline. We compare three forwarding modes: a plain IP forwarding implementation (plain), P4-Protect for unprotected traffic (unprotected), and P4-Protect for protected traffic (protected).

\begin{table}[ht]
\centering
\begin{tabular}[t]{c|ccc}
       & Plain & Unprotected & Protected \\ \hline
    PTI & 100\% & 127\% & 166\% \\
    PTE & 100\% & 127\% & 127\%
\end{tabular}
\caption{Ingress-to-egress packet processing time at PTI and PTE for three forwarding modes: plain, unprotected, and protected.}
\label{table:processing_time}
\end{table}%

%\figeps[0.95\columnwidth]{processing_time}{Ingress-to-egress packet processing time at PTI and PTE for three forwarding modes: plain, unprotected, and protected.}

Table \ref{table:processing_time} shows the ingress-to-egress packet processing time on both PTI and PTE for the three mentioned forwarding modes. The duration is given relative to the processing time for plain forwarding mode. We observe the lowest processing time at PTI and PTE for plain forwarding as it has the least complex pipeline.
With P4-Protect, the processing time at both PTI and PTE is larger than with plain forwarding as the operations are more complex. At PTI, the processing time is even larger with protected forwarding (166\%) than with unprotected forwarding (127\%). 
%This apparently results from pushing the protection header and the IP header for protected traffic. 
At PTE, the processing times for protected and unprotected traffic are equal and 27\% longer than with plain forwarding.

In our implementations, we have used only a minimal IPv4 stack for all three forwarding modes. With a more comprehensive IPv4 stack, the relative overhead through P4-Protect is likely to be smaller.

\subsection{TCP Goodput}
We set up iperf3 connections between client/server pairs and measure their goodput. Each iperf3 connection consists of  15 parallel TCP flows. 
%We utilize the testbed in \fig{tcp_throughput_setup}. 
Two switches are bidirectionally connected via two 100 Gb/s interfaces.
Four client/server pairs are connected to the switches via 100 Gb/s interfaces. Up to 4 clients download traffic from their servers via the trunk between the switches.
% ich habe unterschlagen, dass die clients auf VMs laufen

%\figeps[0.95\columnwidth]{tcp_throughput_setup}{Testbed used for goodput analysis.}

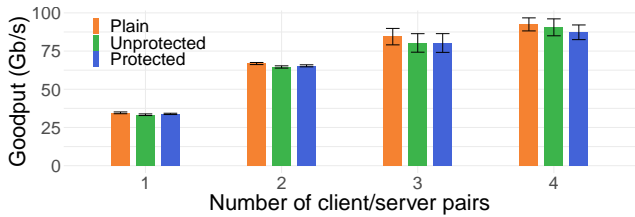
\figeps[0.95\columnwidth]{tcp_throughput}{Impact of varying number of client/server pairs exchanging traffic with iperf3.}

%\twosubfigeps{tcp_throughput_setup}{Testbed: all lines run at 100 Gb/s.}{tcp_throughput}{Average TCP goodput.}{Impact of a varying number of client/server pairs exchaning traffic with iperf3.}

\fig{tcp_throughput} shows the overall goodput for a various number of client/server pairs, each transmitting traffic over a single TCP connection. The goodput is given for the forwarding modes plain, unprotected, and protected.
We performed 20 runs per experiment and provide the 95\% confidence interval.

A single, two, and, three TCP connections cannot generate sufficient traffic to fill the 100 Gb/s bottleneck link. However, with four TCP connections a goodput of around 90 Gb/s is achieved. This is less than 100\% because of overhead due to Ethernet, IP, and TCP headers and due to the inability of TCP to efficiently utilize available capacity at high data rates. Most important is the observation that all three forwarding modes lead to almost identical goodput. The goodput for protected and unprotected forwarding is slightly lower than plain forwarding, which is apparently due to the operational overhead of P4-Protect.

% With a single flow, a host is capable of generating about 35 Gb/s. If two clients are used, a total transmission rate of about 67 Gb/s can be achieved, with three clients about 84 Gb/s and with four clients about 90 Gb/s.

\subsection{Impact on Jitter}

We examine the effect of 1+1 path protection on jitter. 
Two hosts are connected to two Tofino Edgecore Wedges 100BF-32X. The switches are connected with each other via two paths with intermediate Linux servers. Their interfaces are bridged and cause an artificial, adjustable, uniformly distributed jitter. We leverage the \textit{tc} tool for this purpose \cite{LinuxTC}. All lines have a capacity of 100 Gb/s.

%\figeps[0.95\columnwidth]{jitter_setup}{Latency analysis with jitter testbed.}

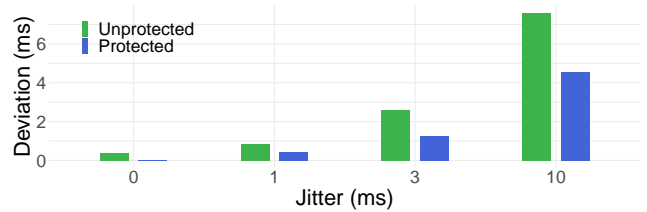
\figeps[0.95\columnwidth]{rtt_jitter}{Impact of 1+1 protection on jitter.}

%\twosubfigeps{jitter_setup}{Testbed: all lines run at 100 Gb/s.}{rtt_jitter}{Average rtt deviation.}{Impact of 1+1 protection on jitter.}

In our experiment, we send pings between the two hosts with and without P4-Protect. \fig{rtt_jitter} reports the average round trip time (RTT) deviation for the pings. Unprotected traffic suffers from all the jitter induced on a single path. Protected traffic suffers only from about half the jitter. This is because P4-Protect forwards the earliest received packet copy and minimizes packet delay occurred on both links.

\section{Conclusion}\label{sec:conclusion}

In this paper we proposed P4-Protect for 1+1 path protection with P4. It may be utilized to protect traffic via two largely disjoint paths. 
We presented an implementation for the software switch BMv2 and the hardware switch Tofino Edgecore Wedge 100Bf-32X. The evaluation of P4-Protect on the hardware switch revealed that P4-Protect increases packet processing times only little, that high throughput can be achieved with P4-Protect, and that jitter is reduced by P4-Protect when traffic is carried over two path with similar delay but large jitter.

%\begin{itemize}
%    \item Simple, high-performance (?) 1+1 path protection scheme
%    \item Implementations for Tofino ASIC and bmv2 simple switch
%    \item Applicable to all kind of traffic
%    \item Applicable beyond domain boundaries as long as PS and PE nodes are under the same control
%\end{itemize}

\bibliographystyle{IEEEtran}
\bibliography{bibs/Local_short}

\end{document}